\documentclass[longbibliography,aps,superscriptaddress,floatfix,letterpaper,preprint]{revtex4-1}

\usepackage{amsmath}
\usepackage{amssymb}
\usepackage{graphicx}
\usepackage[urlcolor=blue, hyperindex, colorlinks, bookmarks=true]{hyperref}
\usepackage{natbib}
\usepackage{mathtools}
\usepackage{times}
\usepackage{amsfonts}
\usepackage{graphicx}
\usepackage{caption}
\usepackage[
top    = 2.75cm,
bottom = 2.75cm,
left   = 2.75cm,
right  = 2.75cm]{geometry}

\begin{document} 

\title{Observation of a Dissipation-Induced Classical to Quantum Transition} 
\date{March 25, 2013}

\author{J. Raftery}\thanks{These authors contributed to the work equally.}
\affiliation{Department of Electrical Engineering, Princeton University, Princeton, NJ 08544, USA}
\author{D. Sadri}\thanks{These authors contributed to the work equally.}
\affiliation{Department of Electrical Engineering, Princeton University, Princeton, NJ 08544, USA}
\author{S. Schmidt}
\affiliation{Institute for Theoretical Physics, ETH Zurich.8093 Zurich, Switzerland}
\author{H. E. T\"{u}reci}
\affiliation{Department of Electrical Engineering, Princeton University, Princeton, NJ 08544, USA}
\author{A. A. Houck}
\affiliation{Department of Electrical Engineering, Princeton University, Princeton, NJ 08544, USA}

\baselineskip24pt

\begin{abstract}
We report here the experimental observation of a dynamical quantum phase transition in a strongly interacting open photonic system. The system studied, comprising a Jaynes-Cummings dimer realized on a superconducting circuit platform, exhibits a dissipation driven localization transition.
Signatures of the transition in the homodyne signal and photon number reveal this transition to be from a regime of classical oscillations into a macroscopically self-trapped state manifesting revivals, a fundamentally quantum phenomenon.
This experiment also demonstrates a small-scale realization of a new class of quantum simulator, whose well controlled coherent and dissipative dynamics is suited to the study of quantum many-body phenomena out of equilibrium.
\end{abstract}

\maketitle 

\clearpage

An understanding of the physics of systems far from equilibrium \cite{national2007} encompasses
deep issues of fundamental importance such as dissipation, decoherence, emergence of classicality from intrinsically quantum systems \cite{RevModPhys.75.715}, symmetry breaking and bifurcations, and how equilibrium is itself established \cite{RevModPhys.83.863,Rigol2008,Trotzky2012,Langen2013}. Unraveling this intricate physics is essential to making sense of the world around us, which is fundamentally nonequilibrium and yet displays complex emergent structure. Much of the important recent progress in experimental condensed matter physics has explored the equilibrium regime of strongly correlated synthetic matter (e.g. ultra-cold atoms in optical lattices \cite{LewensteinBook}), but it has been a long-standing goal to understand what new phenomena may arise as these systems are pushed away from equilibrium. With the rapid technological advances in solid state quantum optics \cite{Blais2004,Wallraff2004}, it is now becoming possible to experimentally study strongly correlated photons, and to build model systems whose open nature gives rise to rich emergent behavior. Interaction with an environment has been argued to provide a mechanism for the emergence of classical behavior \cite{RevModPhys.75.715} from a quantum system. It is also possible, as our work explicitly demonstrates, that dissipation into an environment can qualitatively change this picture, where initially classical dynamics crosses over into one which is fundamentally quantum in nature.

Linear Josephson oscillations \cite{Josephson1962251,RevModPhys.51.101,Sukhatme,PhysRevLett.57.3164,Pereverzev1997} and their anharmonic generalizations when inter-particle interactions are relevant \cite{metastablepistate,PhysRevLett.79.4950,PhysRevA.59.620}, have been observed for atomic Bose-Einstein condensates \cite{Levy2007} (BEC's) and more recently in a system of exciton-polaritons \cite{PhysRevLett.105.120403}. At high density in such BEC systems the large interactions dominate the tunneling and lead to macroscopic quantum self-trapping \cite{PhysRevLett.95.010402,Abbarchi2013}.

In this experiment we explore a localization transition in a dissipative photonic
system \cite{PhysRevB.82.100507}
realized in the circuit quantum electrodynamics (cQED) architecture
\cite{Blais2004,Wallraff2004},
a solid state realization of cavity QED \cite{haroche2006exploring}.
As a system supporting phase-coherent photonic states and controlled nonlinearity
(tunable in situ on nanosecond timescales)
reaching well into the strong-coupling regime even at the single photon level, it opens up the possibility of experimental condensed matter physics with strongly correlated photons.
The flexibility in engineering model Hamiltonians and environmental couplings
makes it an exemplary candidate for carrying out certain classes of quantum simulations
\cite{Feynman1982,Lloyd23081996}
of important but difficult to study problems
\cite{Houck2012,ANDP:ANDP201200261,Greentree2006,Hartmann2006,PhysRevA.76.031805,Koch2009}.
The dynamics of polaritons in driven dissipative Jaynes-Cummings chains have been studied theoretically,
where a transition from classical to non-classical steady state fields,
with varying interaction, tunneling and drive strengths,
observable in the density-density correlation functions, have been suggested
\cite{1367-2630-12-9-093031,PhysRevLett.108.233603}
\footnote{A variant of the previously referenced circuit QED based architectures using networks of
non-linear resonators is presented in
\cite{1367-2630-14-7-075024}}.

The physics of a single qubit coupled to a superconducting microwave resonator is
well described by the Jaynes-Cummings Hamiltonian (we choose units where
$2 \pi \hbar=1$)
\begin{equation} \label{JC-ham}
  \hat{H}^{JC}\:=\:
  \nu_c \ \hat{a}^\dagger \hat{a} \:+\: \nu_a \ \hat{\sigma}^+ \hat{\sigma}^- \:+\:
  g \left(
  \hat{\sigma}^+ \hat{a} \:+\:
  \hat{\sigma}^- \hat{a}^\dagger
  \right) \ ,
\end{equation}
with $\nu_c$ ($\nu_a$) the bare cavity (qubit) frequency and $g$ the qubit-cavity coupling rate, $\hat{a}, \hat{a}^\dagger$ representing the photon annihilation and creation operators, and $\hat{\sigma}^\pm$
the Pauli pseudo-spin operators. The photon-qubit interaction induces an anharmonicity in the spectrum of the Jaynes-Cummings Hamiltonian which leads to an effective on-site repulsion for photons \cite{PhysRevA.76.031805}. Multiple Jaynes-Cummings sites can be coupled to form a lattice with various symmetries and topologies
\cite{Houck2012,ANDP:ANDP201200261,PhysRevA.82.043811,PhysRevA.82.052311,PhysRevA.86.023837}.
Here we study the smallest nontrivial chain, coupling a pair of identical Jaynes-Cummings sites through a photon hopping term (with rate $J$, and subscript $s=L/R$ specifying the left and right sites) to form a dimer
\cite{PhysRevB.82.100507}
\begin{equation}
\label{dimer-ham}
  \hat{H}_{dimer}\:=\:
  \sum_{s=L/R}
  \hat{H}^{JC}_s
  \:-\:
  J \left(
  \hat{a}^\dagger_L \hat{a}_R \:+\:
  \hat{a}^\dagger_R \hat{a}_L
  \right) \ .
\end{equation}
Interaction with the environment is described through a Markovian Lindblad master equation governing the dynamics of the reduced density matrix of the dimer
\begin{equation}
\label{master-eqn}
  \frac{\partial \hat{\rho}}{\partial t} \:=\:
  i \ [ \hat{\rho} , \hat{H}_{dimer} ] \:+\:
  \sum_{i=L,R}
  \Big(
  \frac{\kappa}{2} \: \mathcal{L} [ \hat{a}_i ]
  \Big) \:+\:
  \Big(
  \frac{\gamma}{2} \: \mathcal{L} [ \hat{\sigma}^-_i ]
  \Big) \ ,
\end{equation}
where the Liouvillian super-operator
$
\mathcal{L} [\hat{O}] =
  2 \, \hat{O} \hat{\rho} \hat{O}^\dagger -
  \hat{O}^\dagger \hat{O} \hat{\rho} -  
  \hat{\rho} \hat{O}^\dagger \hat{O}
$
describes the cavity photon and qubit relaxation rates at $\kappa$ and $\gamma$ respectively. Dephasing for our choice of qubits (transmons) can be made much weaker than the above two channels \cite{Koch2007,Schreier2008}, and hence is ignored in our theoretical description.

We first discuss the semiclassical dynamics of the dimer in the absence of dissipation ($\kappa=\gamma=0$). This can be done via the Heisenberg equations of motion and fully factorizing expectation values of spin-photon operator products, yielding a set of eight coupled differential equations for expectation values of the qubit and cavity field operators. A useful representation is in terms of real and imaginary parts of the cavity field $R_s=Re\langle \hat{a}_s \rangle$,
$I_s=Im\langle \hat{a}_s \rangle$,
with the angles parametrizing the spin direction $\vec{n}$ on the qubit Bloch sphere,
$\vec{n}_s=(\sin(\theta_s) \cos(\phi_s),\sin(\theta_s) \sin(\phi_s),\cos(\theta_s))$,
these equations are
\begin{eqnarray} \label{classical-eoms-1}
  \dot{R}_s &=& -\frac{g}{2} \sin(\theta_s) \sin(\phi_s) - J I_{\bar{s}} \nonumber \\
  \dot{I}_s &=& -\frac{g}{2} \sin(\theta_s) \cos(\phi_s) + J R_{\bar{s}}
\end{eqnarray}
for the dynamics of the cavity ($\bar{s}$ denotes the cavity opposite to $s$), and
\begin{eqnarray} \label{classical-eoms-2}
  \dot{\phi}_s &=& -2g \left( R_s \cos(\phi_s) - I_s \sin(\phi_s) \right) \cot(\theta_s) \nonumber \\
  \dot{\theta}_s &=& -2g \left( R_s \sin(\phi_s) + I_s \cos(\phi_s) \right)
\end{eqnarray}
for the qubits.
In writing these equations we have assumed the qubits to be resonant with the respective cavity modes they are coupled to ($\nu_a=\nu_c$), and work in the rotating frame. We define the photon number on the left and right as $N_{L/R}=\langle \hat{a}^\dagger_{L/R} \: \hat{a}_{L/R} \rangle$,
the total photon number as their sum $N=N_L+N_R$,
and the photon imbalance as $Z = (N_L-N_R)/N$.
For special choices of initial conditions, the dynamics can be restricted to certain sub-manifolds of the phase space. One possible choice, $I_1=R_2=0$ and $\phi_1=\pi/2,\phi_2=0$, leads to a set of
of four coupled equations, which preserve this choice.
This sub-manifold contains the dynamics corresponding to an initial condition
with perfect imbalance (e.g. $Z=1$ for $R_1=\sqrt{N}$ and $I_2=0$ at $t=0$).

In the absence of qubit-cavity interaction ($g=0$), the reduced set of equations can be solved exactly, giving rise to harmonic coherent Josephson oscillations of the imbalance at frequency $\nu_J=2J$. With increasing coupling $g$, the oscillations become anharmonic. Solving the system of differential equations numerically (subject to the initial condition with $Z=1$) shows that at a classical critical coupling \footnote{The square root dependence of the critical coupling
\eqref{critical_coupling}
on the photon number
is a consequence of the Jaynes-Cummings nonlinearity \cite{Fink2008}.
At small photon numbers the Jaynes-Cummings ladder is strongly nonlinear, favoring localization.
}
\begin{equation} \label{critical_coupling}
  g_c^{cl} \approx 2.8 J \sqrt{N} \ ,
\end{equation}
the oscillation period diverges, exhibiting critical slowing down,
and resulting in a sharp crossover between two qualitatively different regimes of classical dynamical behavior \cite{PhysRevB.82.100507}, signaling a dynamical phase transition. For couplings beyond the critical value, the system localizes, with the initial photons trapped nearly entirely on a single site, spontaneously breaking the left/right symmetry.
As the parameters $g$ and $J$ are fixed for a particular device, it is helpful to recast the problem in terms of a corresponding classical critical photon number $N_c^{cl} \approx 0.13 \, (g/J)^2$ for a given $g/J$. In the classical analysis,
a dimer initialized with a photon number $N<N_c^{cl}$ is expected to remain in the localized regime
(noting also that the numerical prefactor determining the critical photon number is itself somewhat sensitive to the initial state.)

We now discuss the full quantum dynamics of the dimer in the absence of dissipation ($\gamma=\kappa=0$).
High quality microwave generators acting as classical coherent sources prepare coherent states having nonzero homodyne voltages, making it possible to monitor the system by observing the homodyne quadratures $\hat{I}=(1/2)(\hat{a}+\hat{a}^\dagger)$ and $\hat{Q}=(i/2)(\hat{a}^\dagger-\hat{a})$ (throughout this paper we define the homodyne signal as $\xi=\langle\hat{I}\rangle^2 + \langle\hat{Q}\rangle^2$, whereas the photon number is arrived at by averaging after squaring the individual quadratures, i.e. $\langle\hat{I}^2 + \hat{Q}^2\rangle$. Note that the variables appearing in the classical equations of motion
\eqref{classical-eoms-1} and \eqref{classical-eoms-2}
are the expectation values of the quadrature operators). In the the limit $g\to0$ with finite $J$, initializing the system with a coherent state leads to oscillations of coherent states between the two cavities with a fixed phase difference of $\pi/2$. The oscillations here closely match the expected classical behavior of two coupled oscillators. Keeping $g$ finite and taking $J\to0$, the two Jaynes-Cummings sites decouple, leading to the well-known resonant collapse and revival phenomenon for a coherent state interacting with a single qubit \cite{haroche2006exploring}. From the point of view of the cavity, collapse and revival is a manifestation of the formation of a Schr\"odinger cat state, as each component of the cat state accumulates a different phase due to the interaction with the qubit \cite{PhysRevLett.65.3385,PhysRevA.44.5913}. The use of coherent states emphasizes the stark contrast between the two dynamical regimes - one characterized by classical oscillations and a second by the spontaneous formation of the quintessential macroscopic quantum mechanical state, the Schr\"odinger cat, displaying collapse and quantum revivals.
These two regimes are demarcated by a dynamical quantum phase transition, with the localization
a manifestation of macroscopic quantum self-trapping \cite{PhysRevLett.79.4950}.
We use here the term dynamical quantum phase transition to describe a situation where a qualitative change occurs in the properties of the excited states as a function of a Hamiltonian parameter (here $g/J$), instead of the ground state as in generic quantum phase transitions. The consequence of such a structural change in excited many-body states is reflected in the dynamics of appropriate observables after a quantum quench.

Inclusion of quantum fluctuations results in a renormalization of the critical coupling to its quantum value $g_c^{qu}$ (and likewise for the critical number to $N_c^{qu}$). In figure (1), we show the numerically calculated quantum dynamics of the homodyne signal $\xi$ for initialization of the left cavity with a coherent state of the photon field of varying initial photon numbers (the qubits start out in the ground state and the right cavity in the vacuum state). We note that for the homodyne signal, while the delocalized regime is characterized by harmonic Josephson oscillations at frequency $\nu_J=2J$ as for the imbalance $Z$, the localized regime is marked by fast collapse-revival oscillations the period of which scales as $t_r = \sqrt{N}/g$. In the localized regime, the tunneling is dynamically suppressed and the dimer behaves like two uncoupled Jaynes-Cummings sites.
The transition region around $N^{qu}_c$ displays multi-scale oscillations. At very small photon numbers, we find two further regimes characterized by the reappearance of tunneling and secondary revivals. The richness of the quantum dynamics in the lower part of the figure is due to the finite nature of the system, namely small $N$ and isolation from the environment.  

Figure (2) displays the time averaged quantum expectation value of the imbalance and its fluctuations as a function of $g$, subject to the initialization described above. With increasing $N$, the transition becomes sharper and appears to asymptote at a $g_c^{qu}$ that is smaller than the classical value $g_c^{cl}$. The precise value of the renormalization of the critical coupling, $g_c^{qu}/g_c^{cl}$, depends on the initial quantum state. The crossover region is dominated by large quantum fluctuations and hence is not amenable to a simple mean field description. A natural question to ask is what asymptotic limit gives the semiclassical result described by a sharp transition at $g_c^{cl}$. Our simulations with larger qubit spin $S$ (not shown here) indicate that the appropriate semiclassical limit is $(S,N) \rightarrow \infty$. 

The above arguments apply however to the conservative
 case for which the dimer is isolated and the dynamics conserves the total excitation number
$\hat{N}_T = \sum_{s = L,R} \hat{\sigma}^+_s \hat{\sigma}^-_s + \hat{a}_s^\dagger \hat{a}_s$. We describe below a dynamical phase transition that is of a different nature and is particular to the dimer connected to transmission lines, as studied in our experimental setup. The dynamics of such an open Jaynes-Cummings dimer described by the Master equation (3) does not conserve the total excitation number. As a consequence of this, the photon number decays exponentially and a system initially prepared in the delocalized regime with $N_i \equiv N(t=0)>N_c^{qu}$ will at a finite time cross the phase boundary and localize, breaking the left/right symmetry, as predicted in \cite{PhysRevB.82.100507, PhysRevB.82.100507}. We note that this is distinct from the scenario described above where the transition occurs as a function of parameters $g/J$ in a system that conserves the number of excitations. This transition also differs from nonequilibrium dynamical transitions in the steady state, e.g. when a drive parameter is varied
\cite{Brennecke16072013, 1367-2630-14-8-085011,PhysRevLett.111.220408,PhysRevLett.110.195301,PhysRevE.83.040101}. Interestingly, dissipation drives the system from classical behavior to quantum behavior, contrary to the standard intuition that dissipation always renders systems more classical (for previous work on a quantum to classical transition in a circuit QED realization of single site Jaynes-Cummings physics in the presence of an effective temperature, see
\cite{PhysRevLett.105.163601}). The transition demonstrated in this work stands in sharp contrast to atomic and polaritonic BECs, for which the low-density dynamics is linear \cite{PhysRevLett.79.4950,PhysRevA.59.620,Julia-Diaz2010}, and where dissipation drives the system into a delocalized classical state \cite{Abbarchi2013} \footnote{Other types of interactions of BEC's with an environment can lead to localization. As an example, measurement induced localization arising from spontaneous emission of photons from cold atoms and its implications for optical lattices have been studied in\cite{PhysRevLett.76.3683,PhysRevA.82.013615,PhysRevA.82.063605}}.

Our experimental cQED realization of the Jaynes-Cummings dimer is presented in figures (3a,b).
Each resonator of frequency $\nu_c = 6.34$ GHz and linewidth $\kappa = 225$ KHz is individually coupled to a transmon qubit \cite{Koch2007,Schreier2008} with strength $g = 190$ MHz, providing a strong effective photon-photon interaction. A coupling capacitor allows photon hopping at rate $J = 8.7$ MHz. These parameters place the classical critical photon number at $N_c^{cl} \approx 62$, and enable the observation of many periods of Josephson oscillations $(J\gg \kappa)$. Crucially, at fixed mean initial photon number in the localized phase, there exists an upper bound for $\kappa$ beyond which the averaged revival signal is lost, and the control afforded over dissipation in this architecture allowed us to place $\kappa$ well below this bound, allowing for a good resolution of the quantum revival oscillations \cite{haroche2006exploring}.

The device is operated in both the linear and nonlinear regimes, tuned via external flux lines $V_{L,R}$. To initialize the system (figure (3c)), flux bias pulses shift both qubits far out of resonance, removing photon-photon interactions and allowing efficient population of the linear dimer
modes when driven by a coherent microwave tone $V_{drive}(t)$ at frequency $\nu_c$
modulated by a sinusoid of frequency $J$. Once initialization is complete, and after a variable time delay $\tau$, the nonlinearity is reintroduced by flux biasing the qubits into resonance (this point is our origin of time $t=0$). The delay allows arranging any desired imbalance (and hence oscillation phase) at the beginning of the experiment. Here the imbalance oscillations cover the full range
$-1 \le Z \le +1$, unlike for BEC's \cite{PhysRevLett.95.010402,Abbarchi2013}.

Calibrating the flux pulses  requires locating bias points leading to minimal photon-photon interactions (this corresponds to the minimum of the qubit energy which gives the smallest resonator Lamb shift) for preparation, as well as resonance, where nonlinearity is largest.
The low-lying spectra at these bias points is presented in figure (4),
together with the associated single-photon nonlinearities.
We developed a characterization technique useful for systems with low dissipation that relies on the
Jaynes-Cummings nonlinearity, which leads to bistability with a sharp transition to a bright state
as an applied continuous microwave tone is swept in power
\cite{Boissonneault2010,Bishop2010,Reed2010}.
The threshold for this transition (above which the bright state behaves linearly) is sensitive to the frequency difference between the uncoupled mode being monitored and the nearest low energy polariton mode, a
useful proxy for the strength of the induced nonlinearity.
Such a mapping of the two-dimensional qubit flux space identified
the double minimum and resonance points (see supplementary material).

Dynamics was observed by monitoring photons escaping one of the cavities. After amplification, the signal was mixed down with a local oscillator at $\nu_c$ to produce the $ \hat{I}$ and $\hat{Q}$ quadratures, which were each sampled at $1$ Gs/s. Ensemble averaging over many trials (typically $10^8$) produced the  homodyne signal and photon number (defined previously in terms of the individual quadratures).

If initialized with $N_i<N_c$ ($N_c$ is taken to be the critical photon number observed in the
experiment) the system localizes as soon as interactions are introduced, clearly demonstrated by strong collapse and revival of the homodyne signal \cite{PhysRevLett.58.353,Bloch2002}.
These manifest as a series of lobes with peaks at integer multiples of the the revival time $t_r = \sqrt{N}/g$ and zeros at the midpoints between revivals, a uniquely quantum aspect of decoupled single-site Jaynes-Cummings physics.
Observation of clean revivals, together with an estimate of $g = 190$ MHz, provides a metric for the photon number in the cavity, which is then mapped back to the drive power which created the localized state. A similar calibration is done for the amplitude of the homodyne signal. Exploiting the linearity of the system during initialization we calibrate the initial photon number throughout the full range of drive power.
Figure (5a) shows the revival signal for varying initial photon numbers below the critical value, and $\tau$ chosen to create a perfect initial imbalance $Z=-1$, placing all photons in the monitored cavity.
A gap between clear Josephson oscillations (visible for $N_i>20$) and clear revival dynamics (visible for $N_i<13$) corresponds to the critical region where the revival and Josephson oscillation timescales become
comparable.
The inset presents a fit to the observed revival times displaying the expected $\sqrt{N}$ dependence,
which was used in the calibration described above
for both the initial photon number $N_i$ as a function of initial power, and $\xi_1$, the homodyne signal amplitude of a coherent state with mean photon number of one.
Figures (5b,c) exhibit the preparation of an arbitrary imbalance, where the revival time shows a periodic variation along the vertical axis with $\tau$ (period $1/2J$) and a relative shift between the cavities arising from the imbalance. At the values of $\tau$ corresponding to perfect imbalance ($Z=\pm 1$), all photons are trapped in a single cavity, and the absence of any measurable signal at the other site provides strong evidence for lack of photon tunneling.

For initial photon numbers $N_i > N_c$ the system is placed in the delocalized phase.
Figure (6a) compares the dynamics without and with interactions.
The former case displays quintessential linear behavior - 
exponential decay at a rate $\kappa=225$ kHz modulated by oscillations with frequency $2J$,
proceeding well beyond $N_c$ into the noise floor.
Superposed is a typical example of the nonlinear dissipation driven classical to quantum transition,
initially displaying Josephson oscillations enveloped by an exponential decay. The slightly faster effective decay rate ($\kappa^\prime=265$ kHz) can be attributed to small qubit dissipation and dephasing (as verified by simulations) which play a more significant role with the qubits in resonance, and satisfies the condition for strong single photon nonlinearity $g \gg \kappa^\prime$, a regime not accessible to current exciton-polariton BEC's \cite{Abbarchi2013}.
The exponential decay of the oscillations later gives way to a super-exponential drop in homodyne signal, a signature of the crossover from delocalized to localized behavior: photon escape is a stochastic process, and for a given trial the photon number falls below $N_c$ at a random
time, with an average time dependent on the initial photon number. When approaching this point, the Josephson oscillations become nonlinear, exhibiting a critical slowing down \cite{PhysRevB.82.100507, PhysRevB.82.100507}. Oscillations of different trials within an ensemble dephase with respect to each other, and individual trials once localized exhibit very rapid collapse; thus
ensemble averages of $\hat{I}$ and $\hat{Q}$ die out faster than exponentially, and only trials where $N(t) \gg N_c$ continue to contribute to the homodyne signal. Figure (6b) shows the observed homodyne dynamics for various initial photon numbers, revealing the logarithmic dependence of the critical time to reach the transition on the initial photon number
$t_c \sim \frac{1}{\kappa^\prime} \log (N_i/N_c)$.

Unlike the homodyne signal, photon number measurement is insensitive to the coherence of field in the monitored cavity.
Additionally, the dispersion between individual trials arising from critical slowing down does not cause the photon number to decay super-exponentially.
Figure (6c) compares homodyne measurement to photon number for the same initial condition. For short times the two signals match, demonstrating a high degree of coherence within the ensemble. In contrast to the super-exponential decay of the homodyne signal, we see an exponential decay of the photon number.

The homodyne observation maps out a dynamical phase diagram as a function of initial photon number $N_i$ and time, as displayed in figure (6d). An applied drive during an initialization phase where photon-photon interactions are off initiates Josephson oscillations. An un-driven and non-interacting region lasting a constant time $\tau$ follows, continuing the oscillations, at the end of which the nonlinearity is rapidly switched on. For $N_i>N_c$ there exists a delocalized, interacting phase extending for a time scaling logarithmically with $N_i$, before crossing over into a localized regime. These oscillations are the dominant feature visible for initial photon numbers greater than $20$. With an initialization such that $N_i<N_c$ the system immediately localizes as interactions are switched on, maintaining the trial coherence and hence visibility of the collapse and quantum revival in the homodyne signal. These revivals can be seen as the very short timescale features occurring below $N_i=13$.

We find that the experimentally observed critical photon number ($13 \lesssim N_c \lesssim 20$) is at variance with the classical prediction ($N_c^{cl} \approx 62$), and moreover with the full closed quantum dynamics simulation (placing $N_c^{qu} \approx 120$, as seen in figure (1)). Another feature of the experimental data is the regularity of oscillations close to the transition for $N_i \approx N_c$, which stands in contrast to the irregular oscillations in the full quantum dynamics shown in figure (1). To understand whether dissipation is behind these observations, we have also performed quantum trajectory simulations of the open dimer (that effectively solves the master equation (3)). Figure (7) compares quantum trajectory simulations of an initially coherent state at photon number above $N_c^{qu}$, without and with dissipation. The presence of dissipation is found to linearize the inter-cavity oscillations leading to a homodyne signal that exhibits regular harmonic oscillations, as observed in the experiments.
The variance of the observed critical number from the predicted value
might arise from the role of dissipation in the dynamical transition,
fluctuations in the initial state due to imperfections in preparation,
effect of the higher transmon levels which may be important in the early time dynamics
(though the detuning of the higher levels should suppress their contributions),
and the dynamics associated with quenching in the vicinity of a critical point.

We have demonstrated a nonequilibrium localization transition in a strongly correlated open photonic system.
Our deliberate choice of parameters placed the transition in a region at the margin of accessibility to classical
simulation of the quantum dynamics, providing a test of the reliability of this
architecture as a platform for quantum simulation of open systems \cite{0034-4885-75-8-082401}.
We expect that future experiments will build on this work to explore the intricate dynamical behavior
of lattices when dissipation plays a fundamental role, and which lie
beyond what classical simulators
can replicate.

\paragraph*{Aknowledgements}
The authors would like to acknowledge fruitful discussions with Devin Underwood,
Will Shanks, Srikanth Srinivasan, David Huse, Marco Schiro', Hassan Shapourian, Stephan Mandt and Steve Girvin. 

\paragraph*{Funding}
The Princeton work was supported by
The Eric and Wendy Schmidt Transformative Technology Fund,
the US National Science Foundation through the Princeton Center for Complex Materials (DMR-0819860) and CAREER awards (Grant Nos. DMR-0953475 \& DMR-1151810),
the David and Lucile Packard Foundation,
and US Army Research Office grant  W911NF-11-1-0086.
S.S. acknowledges support through an Ambizione grant of the Swiss National Science Foundation.

\clearpage

\bibliography{paper}

\begin{figure}
\label{Figure1}
\includegraphics[width=6in]{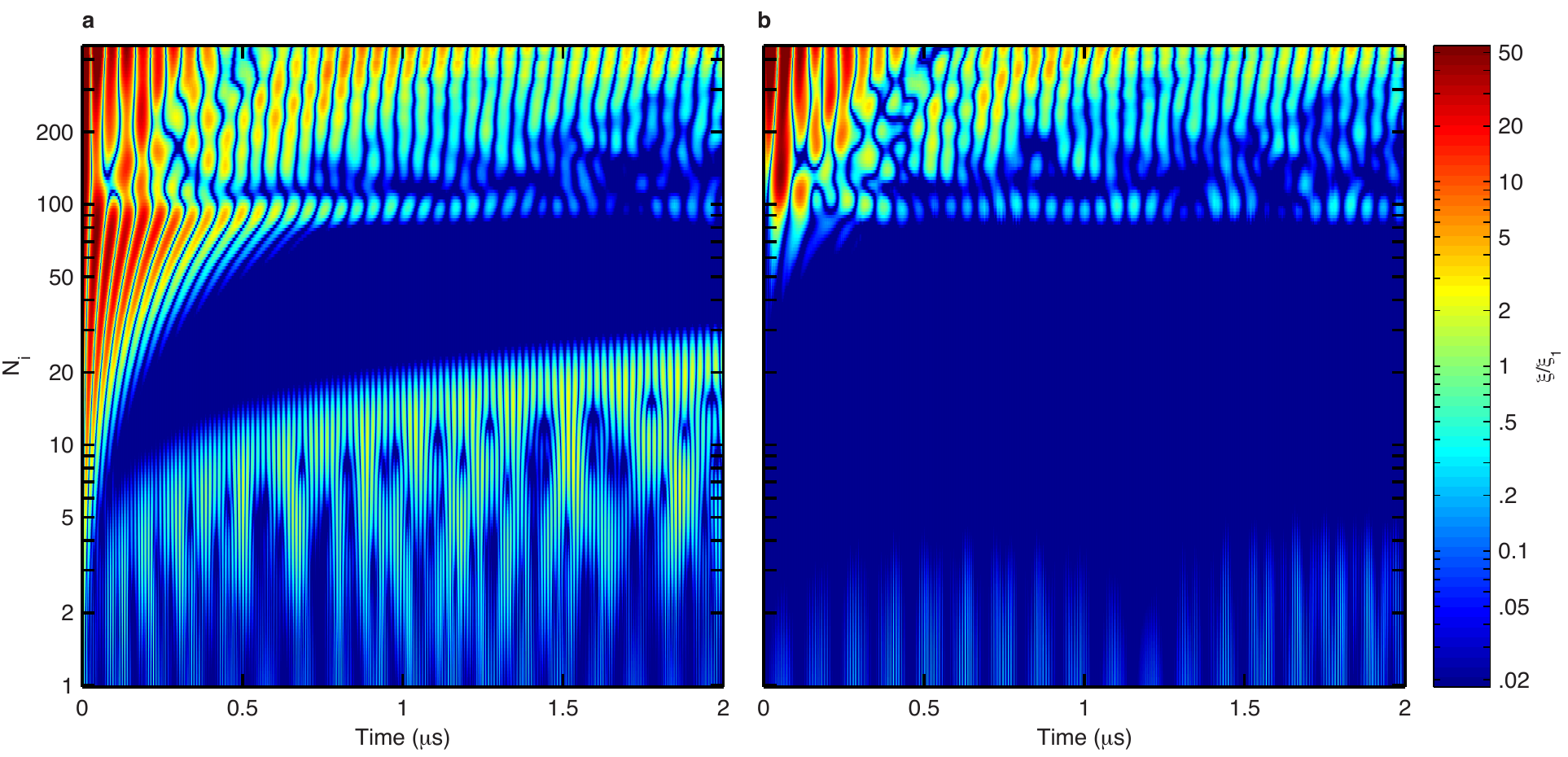}
\captionsetup{justification=raggedright,singlelinecheck=false,font=footnotesize}
\caption{
\textbf{Numerically simulated phase diagram.} Quantum dynamics (without dissipation) of the homodyne signal in left (\textbf{a}) and right (\textbf{b}) cavities, as a function of total photon number (logarithmic vertical axis, with colorbar normalized by $\xi_1$, the homodyne signal of a coherent state with mean occupation of one photon), for the case where the system has been initialized into the $Z=1$ imbalance state, photons in a perfect coherent state, and both qubits
initially in the ground state. The parameters chosen are as given in the main body of the text corresponding to the actual device used in the experiment. For photon numbers below that corresponding to the quantum renormalized critical number $N_c^{qu}$,
tunneling is dynamically suppressed and we observe collapse-revival oscillations with period scaling as $t_r=\sqrt{N}/g$, as expected for decoupled single site Jaynes-Cummings physics \cite{haroche2006exploring}. At very small photon numbers we see an ``exiguous'' regime where tunneling reappears. We also observe in this localized regime secondary revivals at long times (also present in the unitary evolution of the single site Jaynes-Cummings model), which would be washed out in the presence of dissipation. The dynamics above the critical photon number displays Josephson-like oscillations with period $t_J=(2J)^{-1}$, becoming more linear with increasing photon number. We note that the critical number is marked by the coincidence of two time-scales, $t_r=t_J$. As the initial $N$ increases, the revival period grows and ultimately matches that of the Josephson oscillation time-scale, yielding the critical photon number $N_c^{qu} \approx \frac{g^2}{4 J^2} \approx 2 N_c^{cl}$, consistent with where the transition is observed in the simulation. Solving the dynamics at the upper limit shown necessitated the solution of the time-dependent Schr\"odinger equation in a Hilbert space exceeding a dimension of size $10^6$. More details regarding the difficulty of such simulations can be found in the supplemental material.
}
\end{figure}

\clearpage

\begin{figure}
\label{Figure2}
\includegraphics[width=6in]{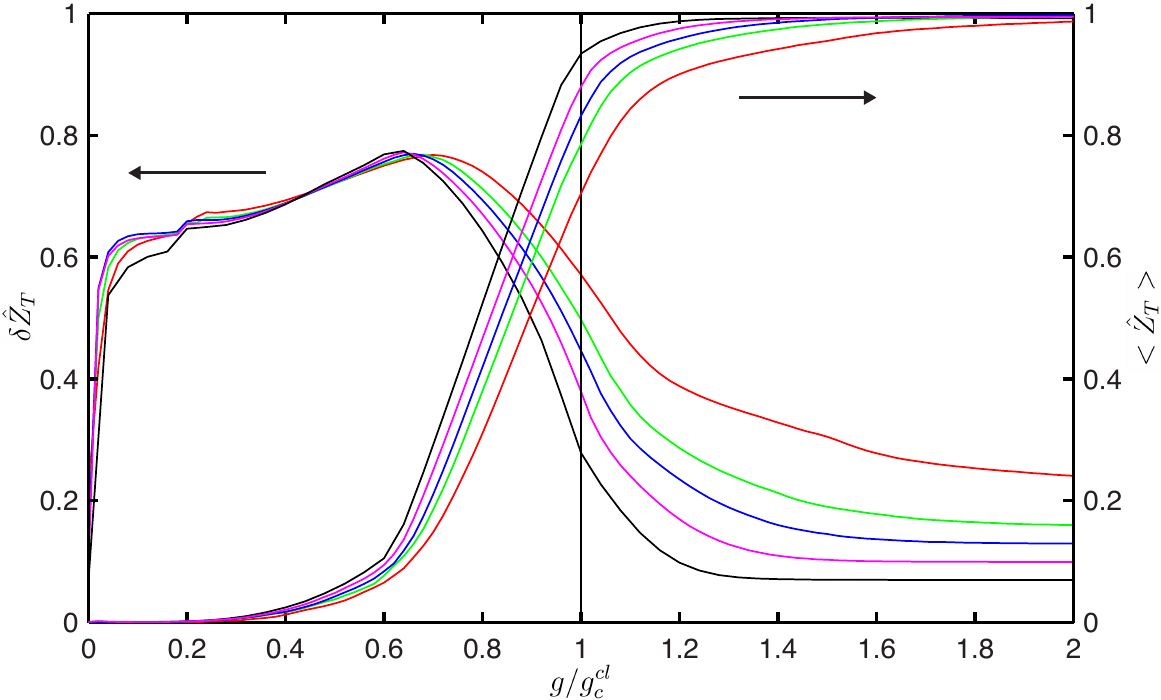}
\captionsetup{justification=raggedright,singlelinecheck=false,font=footnotesize}
\caption{
\textbf{Time averaged imbalance and quantum fluctuations of the imbalance.}
The time averaged imbalance for varying $g/g_c^{cl}$, as well as its quantum fluctuations
(mean squared fluctuations), for coherent
states of three different mean photon numbers
(red $= 20$, green $= 40$, blue $= 60$, magenta $= 100$, and black $= 200$).
Here the total imbalance $Z_T$ is the difference in the number of excitations on each side
(photon plus qubit excitation), normalized by the total excitations $N_T$ in the system.
The classical prediction puts the transition at $g/g_c^{cl}=1$.
The quantum fluctuations are largest in the classically delocalized regime, and lead to a renormalization of
the expected quantum value of the critical coupling $g_c^{qu}$ downwards (i.e. requiring smaller
coupling to observe localization if the total excitation number is held fixed),
and hence relocating the critical photon
number $N_c^{qu}$ upwards.
This is reflected in the buildup of a finite imbalance in the region where the classical analysis
predicts no net imbalance, which is therefore a quantum localized regime.
We also observe that as the number of excitations in the system are increased, the transition gets sharper, suggesting the thermodynamic limit for this spatially finite system is given by the
limit of large excitation number.
}
\end{figure}

\begin{figure}
\label{Figure3}
\includegraphics[width=4in]{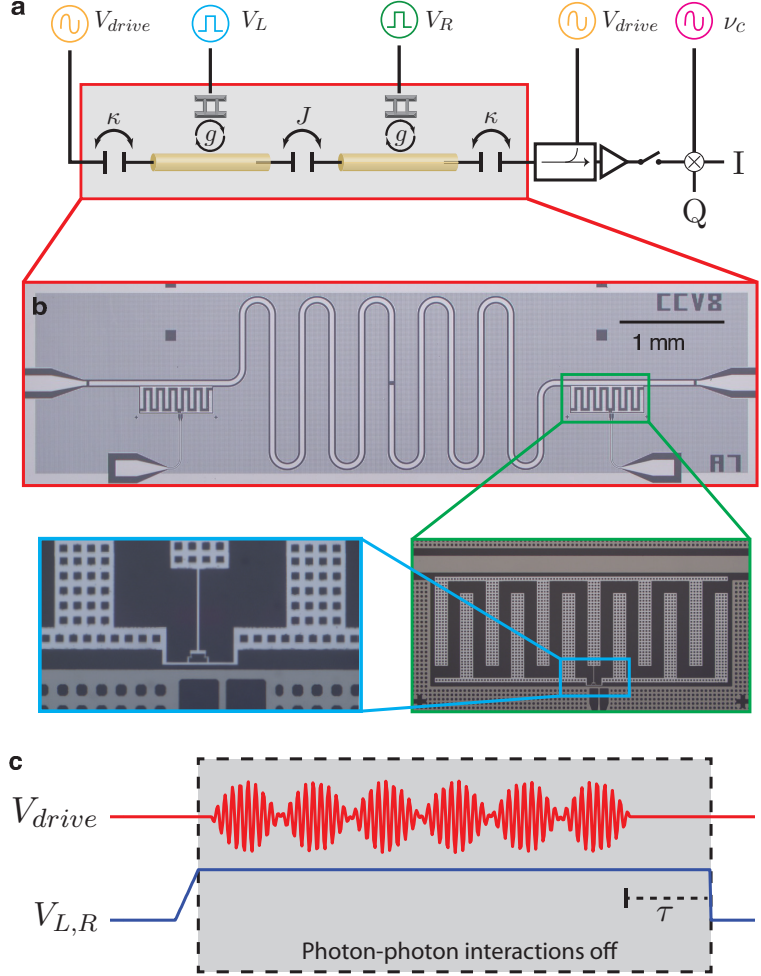}
\captionsetup{justification=raggedright,singlelinecheck=false,font=footnotesize}
\caption{
\textbf{Device layout and initialization routine.}
\textbf{a,} Schematic diagram of the experiment.
The Jaynes-Cummings dimer is comprised of two coupled transmission-line resonators each individually coupled to a transmon qubit.
Inter-cavity coupling $J=8.7$ MHz and cavity-qubit coupling $g=190$ MHz.
The resonators are driven and monitored
via coupling to external transmission lines.
Initialization pulse $V_{drive}$ can be applied to the left or right cavity to generate classical oscillations in the system, while fast flux pulses $V_{L,R}$ control qubit energies at nanosecond timescales.
Right cavity quadratures are monitored via a homodyne measurement.
When $V_{drive}$ is applied to the right cavity a fast microwave switch is used to block the strong reflected signal.
\textbf{b,} Optical micrograph of the device.
\textbf{c,} Initialization routine pulse waveforms.
Fast flux pulses $V_{L,R}$ rapidly detune both qubits to their minimum energies to turn off photon-photon interactions.
While qubits are detuned either the left or right cavity is driven with initialization pulse $V_{drive}=\sin(2 \pi \nu_c t)\sin(2 \pi Jt)\Theta(-t)\Theta(t+\frac{m}{J})$, where $m$ is an integer,
and $\Theta$ is the Heaviside step function.
Variable delay $\tau$ after the end of $V_{drive}$ allows the photon-photon interactions to be turned on at any any point during the un-driven linear oscillations, enabling the preparation of any desired imbalance.
}
\end{figure}

\begin{figure}
\label{Figure4}
\includegraphics[width=3.2in]{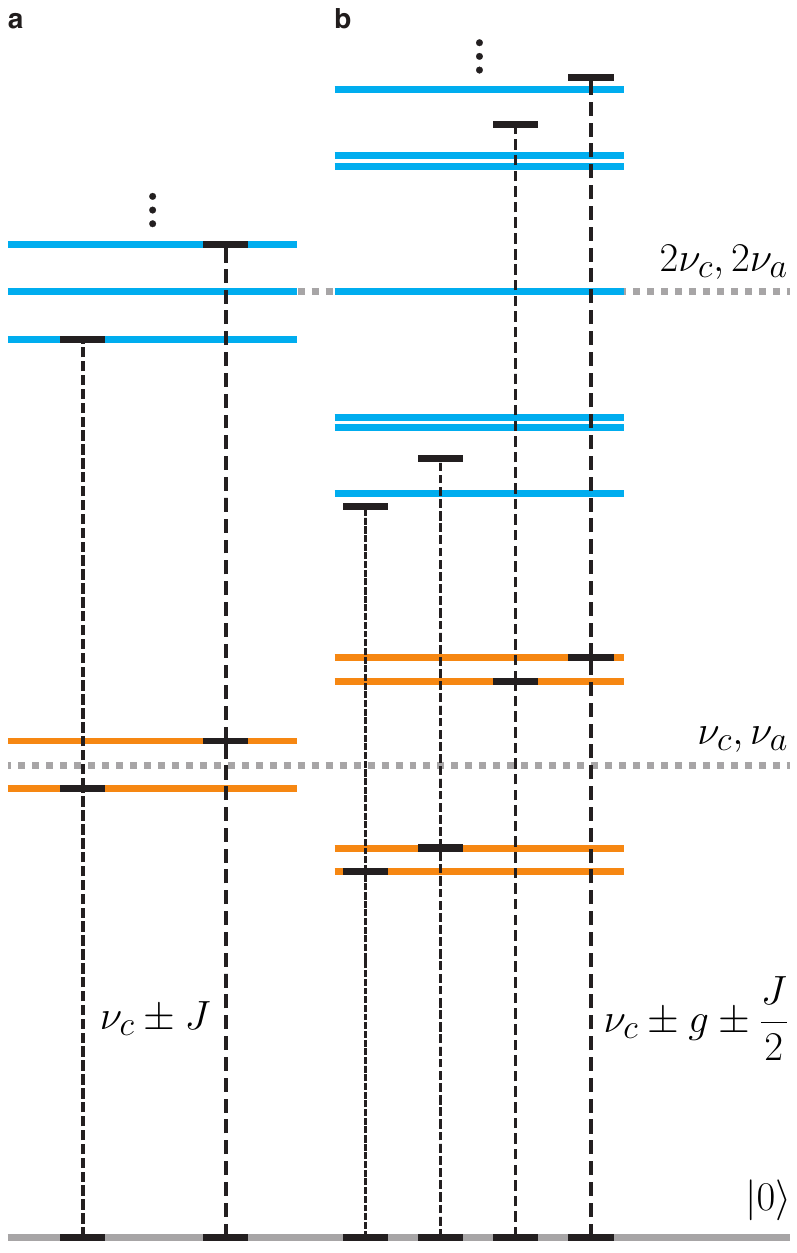}
\captionsetup{justification=raggedright,singlelinecheck=false,font=footnotesize}
\caption{
\textbf{Low-lying dimer spectra and nonlinearities.}
Spectra are shown for the two flux bias points in the experiment, without (a) and with (b) photon-photon interactions.
The first (orange) and second (blue) excitation manifolds are shown along with transitions (black), where transitions from the first to second excitation manifold are the same frequency as the ground to first transitions immediately below.
\textbf{a,} Effective photon-photon interactions can be turned off by tuning qubits to their minimum energies.
While detuned, the qubits remain in their ground states and can be ignored, resulting in a simplified spectrum of two cavities in resonance.
The $J$ coupling creates two linear hybridized modes with energies $\nu_c \pm J$ , ideal for generating full linear oscillations ($Z(t)$ oscillates between $\pm1$ ).
Modulating $V_{drive}$ at $J$ generates sidebands resonant with each mode and explicitly sets the phase of the resultant linear oscillations, generating an imbalanced coherent state at $t=0$.
\textbf{b,}
Photon-photon interactions are generated by tuning both qubits into resonance with the cavities.
At this bias point the first excitation manifold has four polariton states at energies $\nu_c \pm g \pm J/2$.
Strong single photon nonlinearities are apparent, as no transitions to the second manifold match the energies of the first manifold.
Due to the form of the Jaynes-Cummings hamiltonian, all nonlinearities are photon number dependent and lead to linear behavior at high excitation manifolds.
}
\end{figure}

\begin{figure}
\label{Figure5}
\includegraphics[width=6.5in]{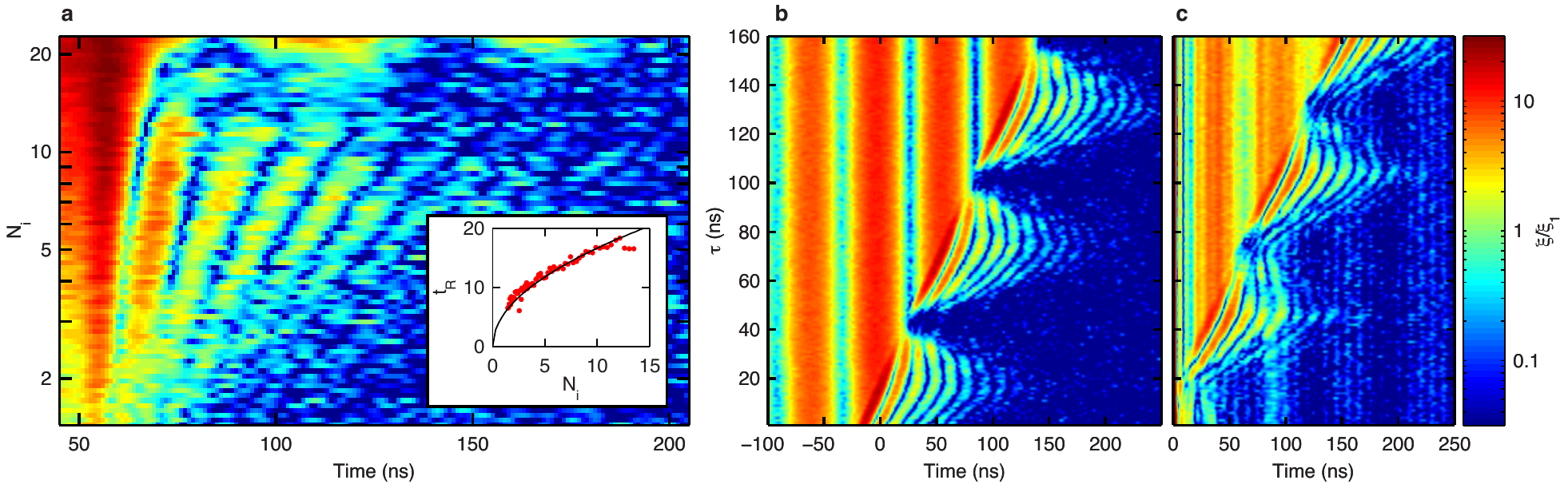}
\captionsetup{justification=raggedright,singlelinecheck=false,font=footnotesize}
\caption{
\textbf{Self-trapped regime. a,} Homodyne signal dynamics of the right cavity as a function of initial photon number and time.
A $345$ ns ($3 / J$) initialization pulse $V_{drive}$ ending at $t=0$ generates linear oscillations between the left and right cavities while photon-photon interactions are off.
The oscillations proceed for a time $\tau$ such that a perfect imbalance ($Z=-1$) is established when interactions are turned on.
For $N_i<N_c$ the system is localized and the right cavity displays fast collapse and revival oscillations while the left cavity (not shown) remains empty.
(Inset)
Fitting revival time $t_r$ as a function of drive power to the expected $\sqrt{N}$ dependence allows drive power to be mapped to initial photon number (at $t=0$), as well as the calibration of $\xi_1$, the homodyne signal of a coherent state with mean occupation of one photon.
\textbf{b-c,}
Revival time variation as a function of $\tau$ for $V_{drive}$ applied to the left (left panel) and right (right panel).
As $\tau$ is varied the full range of imbalance is observed.
Here the same initialization scheme as in (a) is used with $N_i=8$, which produced the cleanest collapse and revival signature.
When driving the right cavity, strong reflections during $V_{drive}$ give rise to signal distortion during the dynamics (visible in the right panel) which are mitigated by using a microwave switch.
}
\end{figure}

\begin{figure}
\label{Figure6}
\includegraphics[width=6.5in]{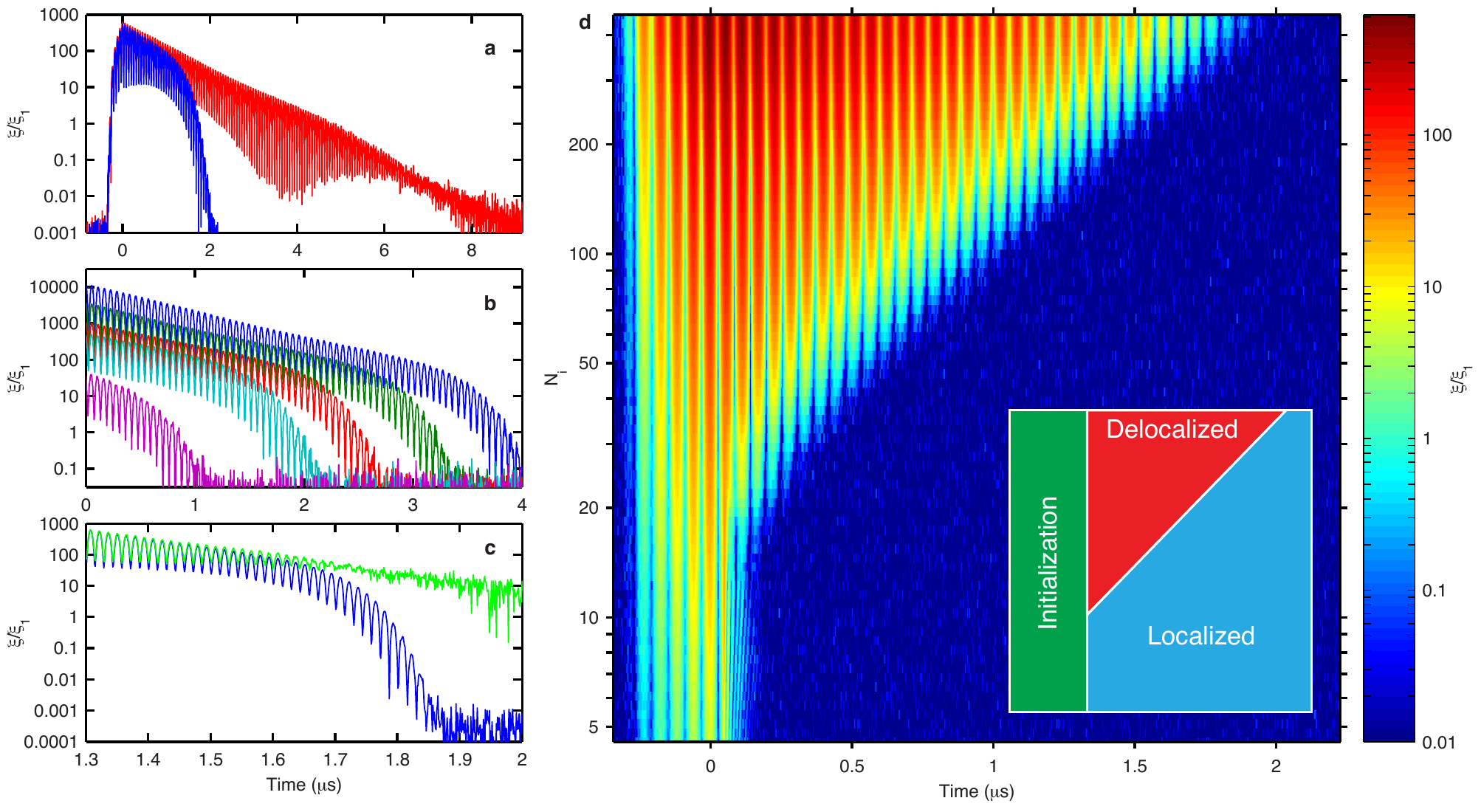}
\captionsetup{justification=raggedright,singlelinecheck=false,font=footnotesize}
\caption{
\textbf{Dissipation driven transition and phase diagram. a,}
Comparison of homodyne signal when photon-photon interactions are off (red) and on (blue) with
$N_{i}>N_c$.
The same initialization pulse $V_{drive}$ was used as in figure (5).
With interactions off, the system undergoes oscillations and exponential decay at rate $\kappa$, which can be observed for several microseconds.
Significantly, the presence of interactions causes super-exponential decay of the homodyne signal as $N$ approaches $N_c$, a signature of crossover into the localized regime.
\textbf{b,}
The time of onset of super-exponential decay, $t_c$, shifts with initial photon number, as shown here for
initialization pulses with varying drive power
lasting $11.5\:\mu s$ ($100/J$).
The use of a long initialization pulse makes it possible to drive the system to very large initial photon numbers (top to bottom $N_i \approx$ 12,000; 3,800; 1,100; 550; 40), but introduces complications (see supplementary material for more details).
\textbf{c,}
Directly measuring photon number (green) reveals that incoherent photons remain in the system after the homodyne signal (blue) has undergone super-exponential decay.
Oscillations in the photon number can also be observed to die out,
as critical slowing down constrains the envelope of oscillations, finally leaving only exponential decay.
Here $V_{drive}$ is $1.15\:\mu s$ ($10 / J$) and $\tau=1\:\mu s$.
Background voltages leading to distortion of the signal were removed from the photon number measurement.
\textbf{d,}
Reconstructing the phase diagram by monitoring the homodyne signal as a function of initial photon number and time. At high powers the dynamical transition from linear oscillations to localized behavior is marked by super-exponential decay, while at low powers the collapse and revival signatures of localized behavior are observed.
A $345$ ns ($3 / J$) initialization pulse $V_{drive}$ ending at $t=0$ was used with $\tau = 65$ ns,
corresponding to an initial imbalance $Z \approx -0.6$.
(Inset) Illustration of the phase diagram showing the different dynmical regimes.
}
\end{figure}

\begin{figure}
\label{Figure7}
\includegraphics[width=6in]{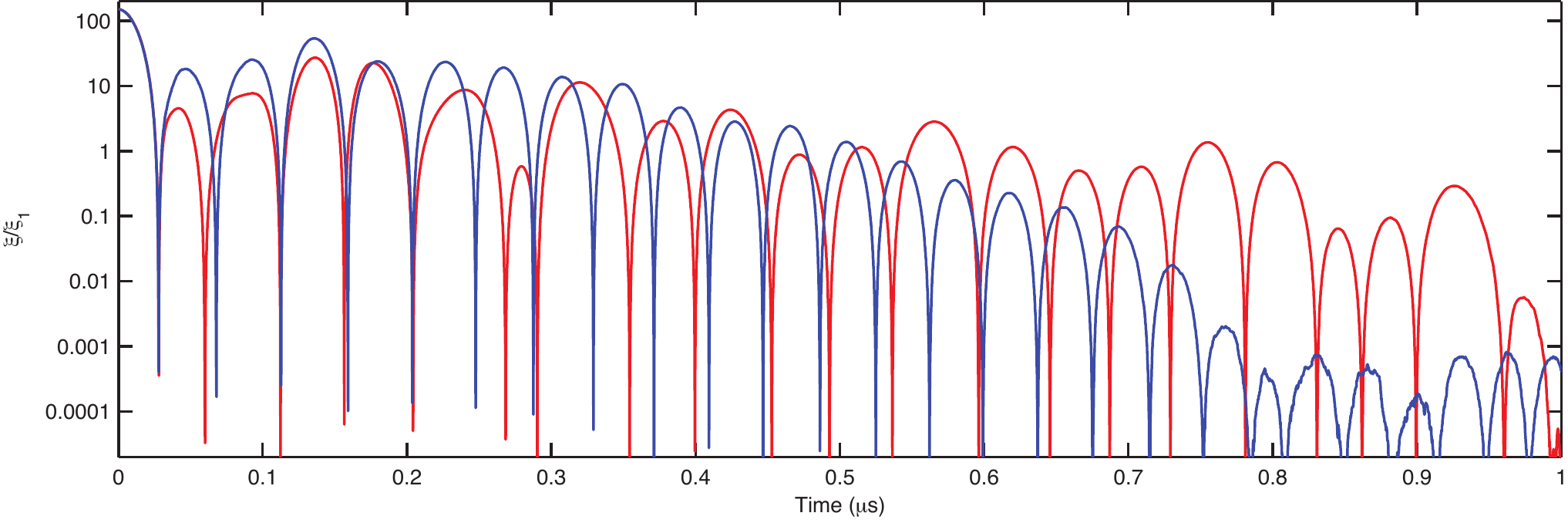}
\captionsetup{justification=raggedright,singlelinecheck=false,font=footnotesize}
\caption{
\textbf{Simulated homodyne signal without and with dissipation.}
A comparison of the homodyne dynamics (shown on a logarithmic scale, normalized to $\xi_1$, the homodyne signal of a single photon coherent state)
without dissipation (red curve), and with (blue curve, averaged
over $1000$ walks in a quantum trajectory simulation of the master equation).
The initial state is a coherent one with a mean of $150$ photons on the left, no photons on the right, and the qubits in the ground state. The unitary dynamics (no dissipation) exhibits nonlinear oscillations due to the competition between inter-cavity tunneling and qubit coupling, in the delocalized regime not very far above $N_c^{qu}$. As the tunneling is less sensitive to dissipative effects, the tendency is for dissipation to linearize the oscillations. The early overshoot in the homodyne signal results from transient behavior in the initial dynamics of the system.
}
\end{figure}

\clearpage

\section*{\large{Supplementary Materials}}

\paragraph*{\textbf{Characterizing the Jaynes-Cummings Dimer}}

Characterizing the system involves locating the two qubit flux bias points required for (1) efficient initialization and (2) strong photon-photon interactions during dynamics.
Since both qubits are individually tunable the system is characterized by two independent voltages $V_L$ and $V_R$.
The conventional method for mapping out circuit-QED systems is to look at the frequency response in transmission at low drive power as a function of flux bias voltages.
This works because the low-lying polariton mode frequencies are sensitive to qubit detunings, and allow direct measurement of the spectra.

Our early physical realizations of dimers with relatively high dissipation rates ($\kappa \approx 10$ MHz) were successfully characterized in this manner.
Figure (S1) shows a map of an older device as a function of $V_L$ and $V_R$.
Here drive frequency is fixed and transmission through the dimer is monitored, producing a large signal only when a first excitation manifold polariton mode with a significant resonator component shifts to be resonant with the drive frequency.
A frequency far from $\nu_c$ is chosen, highlighting regions of flux space near resonance.
The map shows the periodicity due to flux quantization, with the non-orthogonality of lattice vectors stemming from cross-coupling of the flux biases.
The indicated contour on figure (S1) brings both qubits through resonance, one from above ($\nu_a^{L}>\nu_c$) and one from below ($\nu^{R}_a<\nu_c$).
Transmission as a function of drive frequency along this contour is shown in figure (S2).
At resonance, the four polariton modes are clearly visible at the expected frequencies ($\nu_c \pm g \pm J/2$).

Calibration using this method has proven challenging for later devices fabricated with very low dissipation, important for observing long timescale dynamics.
Polariton modes with a significant qubit component (true for all modes in resonance) are particularly difficult to measure, as dissipation channels not contributing to measurement reduce the signal to noise ratio (SNR).
This difficulty is compounded by the presence of multiple cavities, with the end result that low power characterization of low dissipation dimers has proven impractical.
A new technique is required for rapid mapping of flux space.

The bistability due to the Jaynes-Cummings nonlinearity can be used to probe the spectrum quickly and with large SNR.
The technique is similar to the high fidelity qubit state readout technique of Reed et al.
\cite{Reed2010} (see also \cite{Boissonneault2010,Bishop2010}).
In their experiment, a difference in cavity frequency when the qubit is in the ground or excited state ($\nu \approx \nu_c \pm \frac{g^2}{\Delta} \sigma_z$, where $\Delta$ is the detuning of the qubit from the cavity) creates a qubit state dependent threshold power, which is the drive power required to enter the bright state when the cavity is driven by a pulse at the bare resonator frequency.
Setting a pulse power between the two threshold powers leads to high distinguishability between the ground and excited state of the qubit.

For the dimer, we take advantage of the fact that the threshold behavior is sensitively dependent on the proximity of low-lying polariton modes to the linear dimer modes (where qubits are decoupled but the resonator coupling leads to a symmetric ($\nu_S$) and an antisymmetric ($\nu_A$) mode, where $\nu_{S,A}=\nu_c \pm J$, and the lower energy mode is the symmetric one).  
Transmission through the dimer is monitored while a continuous microwave drive ($\nu_d = \nu_{S,A}$) is swept in power.
Very little averaging is required to distinguish the dark and bright state, greatly speeding up the time required to perform a characterization.

Plotting threshold power as a function of the two flux biases produces maps like the one shown in figure (S3), which is for the antisymmetric mode.
Threshold power varies by over two orders of magnitude, making it easy to pick out important features.
The blue regions are immediately noticeable and can be identified with minimum qubit energies leading to incredibly weak nonlinearity.
This weak nonlinearity is ideal for loading photons into the dimer, and is used during initialization.
Additionally, locating these reference points simplifies the search for resonance by restricting the space needed to be searched to a contour in flux space connecting the $(\Phi_L, \Phi_R)=(-1/2,-1/2)$ and $(\Phi_L, \Phi_R)=(1/2,1/2)$, with flux expressed in units of the superconducting flux quantum
$\Phi_0=h/2e$.
Since both qubits are nearly identical, this contour simultaneously moves both qubits from their minimum energies to their maximum energies, passing through resonance together.

This contour is shown in figure (S4) for both the antisymmetric (a) and symmetric (b) modes.
Averaged transmission is shown as a function of power and flux, along a contour in two-dimensional
flux space,
revealing the full structure of the transition from dark state into bright state, beyond just a simple threshold power.
The detailed shape of the transition region depends on the specific settings of the power ramp (sweep time, starting power, IF bandwidth) likely stemming from the effect these settings have on the time spent at different powers and hence different probabilities of transition.
In addition to broad shifts of the threshold power which we attribute to movement of the lowest excitation manifold, a good deal of fine structure is apparent, thought to be due to the movements of the many states in higher excitation manifolds.
When a higher excitation manifold state lines up with $\omega_{S,A}$, multi-photon transitions become important, despite their low probability.
The precise details of this behavior are not fully understood at this time, but will be the subject of
further work.

In the dimer system used throughout the paper, qubit maximum energies were built to be just above resonance with the cavities.
This was important for reducing sensitivity to flux noise, but made locating the exact resonance point challenging.
Resonance was identified as a point of very high nonlinearity (high threshold power) that simultaneously displayed a high degree of symmetry between the transition shape of the symmetric and antisymmetric dimer modes.
For this experiment, a flux bias of  $(\Phi_L, \Phi_R)=(-0.0286,-0.0286)$ was chosen as our resonance bias point.

\paragraph*{\textbf{Toy Model Showing Super-exponential Decay}}

In this section we present a simplified toy model which captures the essence of the super-exponential decay as photon leakage takes the system from a region of Josephson oscillations through the cross-over and into the localized regime, presupposing the existence of the two regimes and their dynamical features.

In this model we average over a large number ($10^6$) of individual trials, each trial accounting for photon leakage probabilistically.
All trials begin with an initial number of photons $N_i$ in the dimer, each of which escapes according to a Poisson process with decay rate $\kappa=225$ kHz.
So long as the photon number at a given time is above the critical number for the transition ($N_c=20$ for this simulation), the system is assumed to undergo linear oscillations at frequency $2J$, with a homodyne signal amplitude that scales with the photon number.
When the photon number for the single trial crosses the critical threshold, the oscillations of that trial are extinguished, setting the amplitude to zero.
The stochastic nature of photon leakage leads to the vanishing of the oscillation at a random time in each trial.
Averaging over all trials in the ensemble produces the expected super-exponential decay signature, as shown for representative initial photon numbers in figure (S5a).
In figure (S5b) we plot the toy model's output as a function of $N_i$ and time, which produces an analog of the phase diagram appearing as figure (4c) in the main text.

In the actual experiment, as the region of the cross-over is approached in a single trial, the oscillations exhibit a critical slowing down, which appears as a randomization of the relative oscillation phase across trials, and hence an averaging (across all trials in an ensemble) to zero of the oscillations.
Additionally, when an individual trial has localized, the quantum collapse and revival leads to a rapid disappearance of the homodyne signal.
When further account is taken, within the toy model, of the critical slowing down of the oscillation frequency, the phase boundary is shifted toward a higher critical photon number.

\paragraph*{\textbf{Driven Dynamics}}

In nearly all experiments shown in this paper the initialization procedure involves a high-power drive applied when the qubits are far detuned.  
This ensures that the system is well above the Jaynes-Cummings threshold for linear system behavior during initialization.  
It is also important to note that the length of the drive pulse (typically 3/J) is also short compared to the time it takes for the system to reach a steady state where drive and dissipation balance.
Keeping the drive short ensures the initial photon number is a linear function of drive power, which is necessary for the calibration using the localized revival time behavior to be applicable at higher drive powers (see the discussion of figure (3a) in the main text).
Unfortunately such short drive pulses limited the maximum initial photon number to about $500$.
To linearly delay the dissipation-induced transition it is necessary to exponentially increase $N_i$, which was accomplished using a much longer initialization procedure in figure (4b) in the main text ($11.5 \mu s$ or $100/J$).

Figure (S6) shows the same data as in figure (4b) but includes the driven dynamics observed during the initialization period before photon-photon interactions were turned on by tuning the qubits into resonance.
For drive powers above the Jaynes-Cummings threshold (-5 dBm) the system behaves as expected: super-exponential decay sets in at a time logarithmically dependent on the initial photon number.
For lower powers, however, the driven dynamics show some interesting features due to the finite nonlinearity of the dispersive spectrum at low photon numbers.
Josephson oscillations that build up over the first few microseconds of drive disappear, followed by a reappearance of the homodyne signal microseconds later.
The full account of the driven-dissipative dynamics is beyond the scope of this work, and will require further study. 

\paragraph*{\textbf{Simulation of Jaynes-Cummings Lattices on Classical Computers}}

We briefly comment here on the difficulty of simulating the physics of Hamiltonians involving
Hilbert spaces built as tensor products of many subsystems.
We recall here the dimer Hamiltonian from the main body of the paper
\begin{equation}
  \label{dimer-ham-supp}
  \hat{H}_{dimer}\:=\:
  \hat{H}^{JC}_L\:+\:
  \hat{H}^{JC}_R\:-\:
  J \left(
  \hat{a}^\dagger_L \hat{a}_R \:+\:
  \hat{a}^\dagger_R \hat{a}_L
  \right) \ ,
\end{equation}
with single site Jaynes-Cummings Hamiltonian being
\begin{equation}
  \hat{H}^{JC}\:=\:
  \nu_c \ \hat{a}^\dagger \hat{a} \:+\: \nu_a \ \hat{\sigma}^+ \hat{\sigma}^- \:+\:
  g \left(
  \hat{\sigma}^+ \hat{a} \:+\:
  \hat{\sigma}^- \hat{a}^\dagger
  \right) \ .
\end{equation}
The dimension of the Hilbert space $\mathcal{H}$ necessary for simulating the Hamiltonian \eqref{dimer-ham-supp}
scales as $dim \: \mathcal{H} \sim (2 \eta)^2$, with $\eta$ the cutoff on the photon Fock space.
For a lattice of $n$ Jaynes-Cummings sites, this becomes $dim \: \mathcal{H} \sim (2 \eta)^n$,
showing an exponential scaling of the Hilbert space size, and making classical simulations
intractable for even quite small systems.
This Hamiltonian conserves total excitation (polariton) number
$\hat{\vartheta} = \sum_{s=L/R} \left( \hat{N}_s+\hat{\sigma}_s^+ \hat{\sigma}_s^- \right)$
($\hat{N}_s$ is the total photon number operator on site $s$, while $\hat{\sigma}_s^+ \hat{\sigma}_s^-$
counts excitations of qubit $s$), i.e. $\left[ \hat{H}_{dimer} , \hat{\vartheta} \right]=0$,
and so $\hat{H}_{dimer}$ is simultaneously diagonalized in an eigenbasis of $\hat{\vartheta}$.
Writing the dimer Hamiltonian \eqref{dimer-ham-supp} in such an eigenbasis requires only
$4 \langle \hat{\vartheta} \rangle$ states.
However, as discussed in the main body of paper, we are interested in following the dynamics of coherent
states of the cavity fields, which are represented as superpositions over Fock states representing
many different photon numbers, and hence such a restriction to fixed polariton number subspaces is
insufficient to capture their dynamics. Inclusion of dissipation removes the conservation law, and
immediately necessitates use of the full truncated Hilbert space.
Additionally, simulating the master equation to capture the dissipative couplings to an environment
via quantum trajectory techniques increases the difficulty of the simulation multiplicatively, by the number
of stochastic walks to be averaged over (of order hundreds to thousands of individual walks).

\clearpage

\begin{figure}
\includegraphics[width=4in]{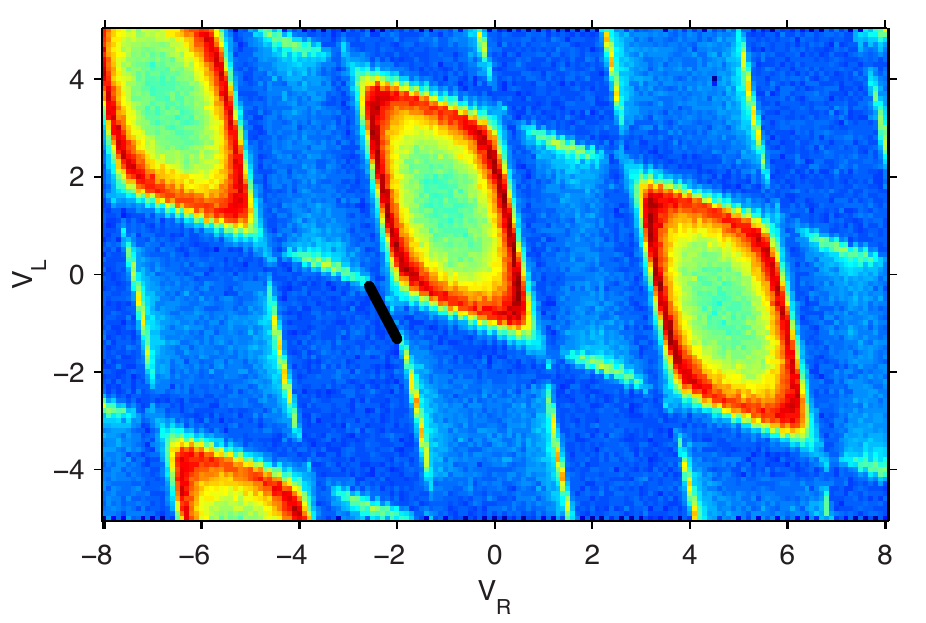}
\captionsetup{justification=raggedright,singlelinecheck=false,font=footnotesize}
\caption{\textbf{S1.}
Transmission through an earlier realization of the Jaynes-Cummings dimer with larger dissipation ($\nu_c =6.182$ GHz, $J=24$ MHz, $\kappa=11$ MHz) as a function of $V_L$ and $V_R$ for a fixed frequency ($\nu_d = 6.131$ GHz).
The black contour indicates the path in flux space followed in figure (S2), which passes through resonance.
}
\end{figure}

\begin{figure}
\includegraphics[width=5in]{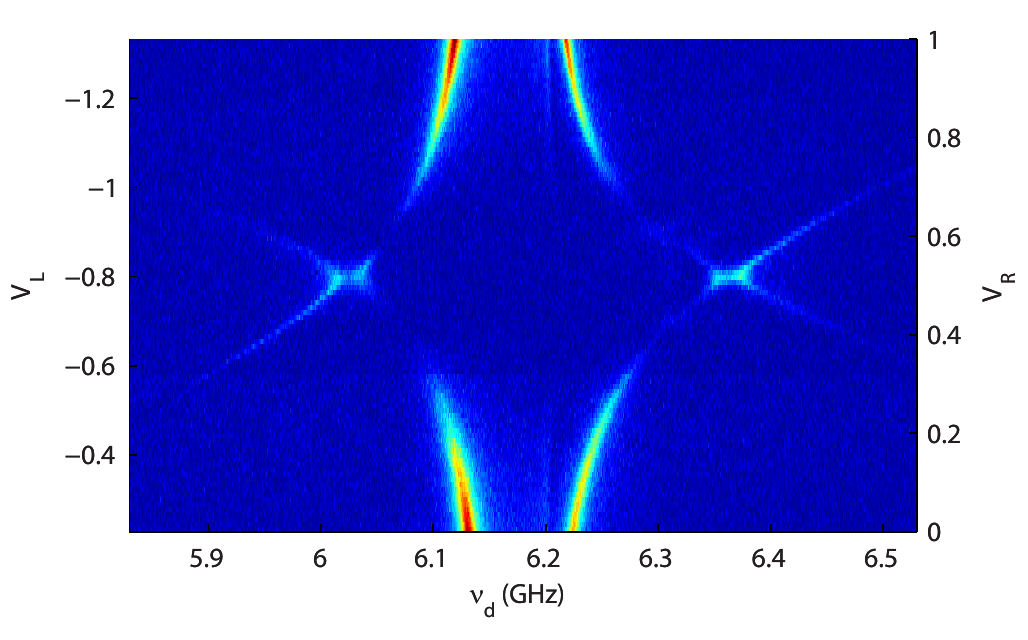}
\captionsetup{justification=raggedright,singlelinecheck=false,font=footnotesize}
\caption{\textbf{S2.}
Low power transmission through an earlier realization of the Jaynes-Cummings dimer with larger dissipation, as a function of flux bias and drive frequency.  The flux contour was chosen so that both qubits simultaneously enter resonance, one from above ($\nu_a^{L}>\nu_c$) and the other from below ($\nu^{R}_a<\nu_c$). 
}
\end{figure}

\begin{figure}
\includegraphics[width=6in]{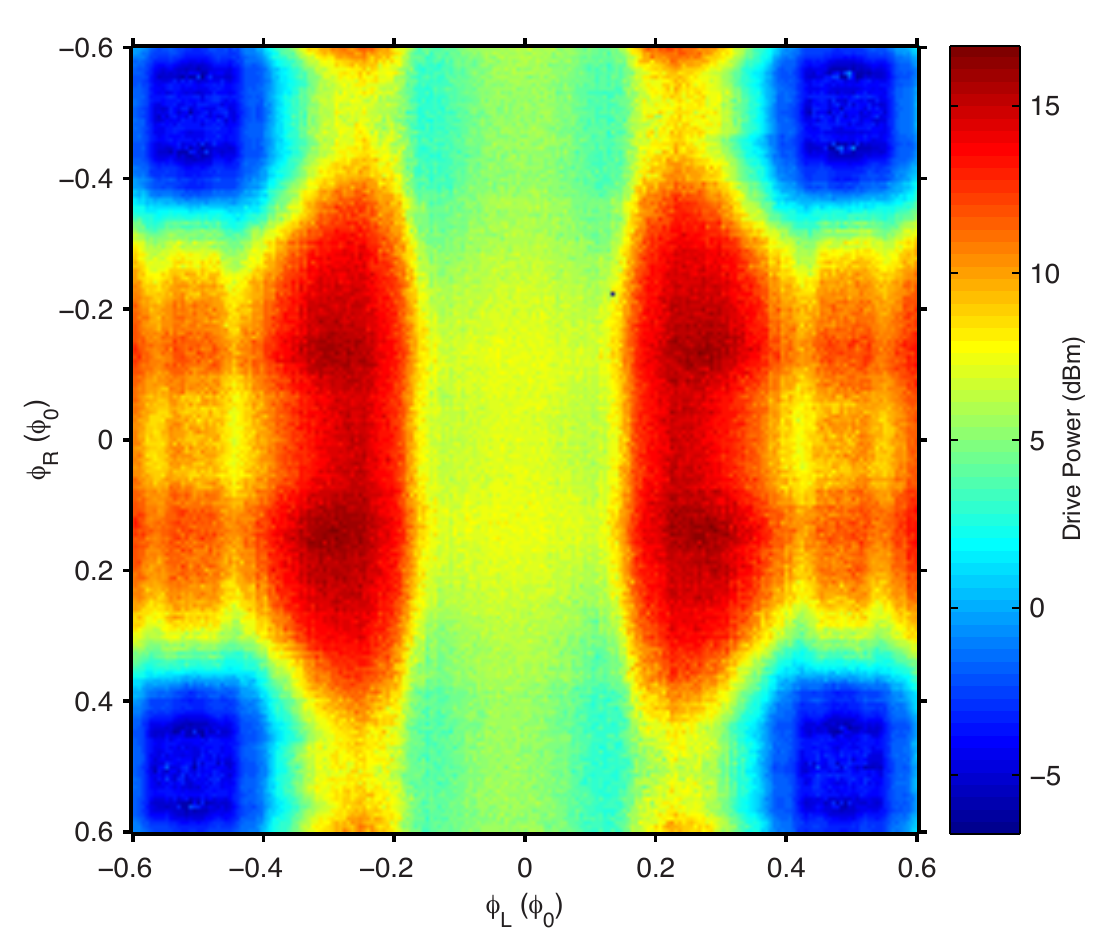}
\captionsetup{justification=raggedright,singlelinecheck=false,font=footnotesize}
\caption{\textbf{S3.}
Threshold power for entering the bright state as a function of left and right qubit flux bias for the asymmetric cavity mode.  
The biases $V_L$ and $V_R$ have been normalized to be expressed in units of the flux quantum $\Phi_0$ and remove cross coupling.
At each coordinate in flux space transmission was measured in response to a microwave tone swept in power from $-20$ dBm to $20$ dBm and resonant with the asymmetric cavity mode.
Only $0.5$ seconds of averaging was needed to obtain a clean signal from which the threshold power was extracted, allowing for rapid calibration of low dissipation systems.
}
\end{figure}

\begin{figure}
\includegraphics[width=6in]{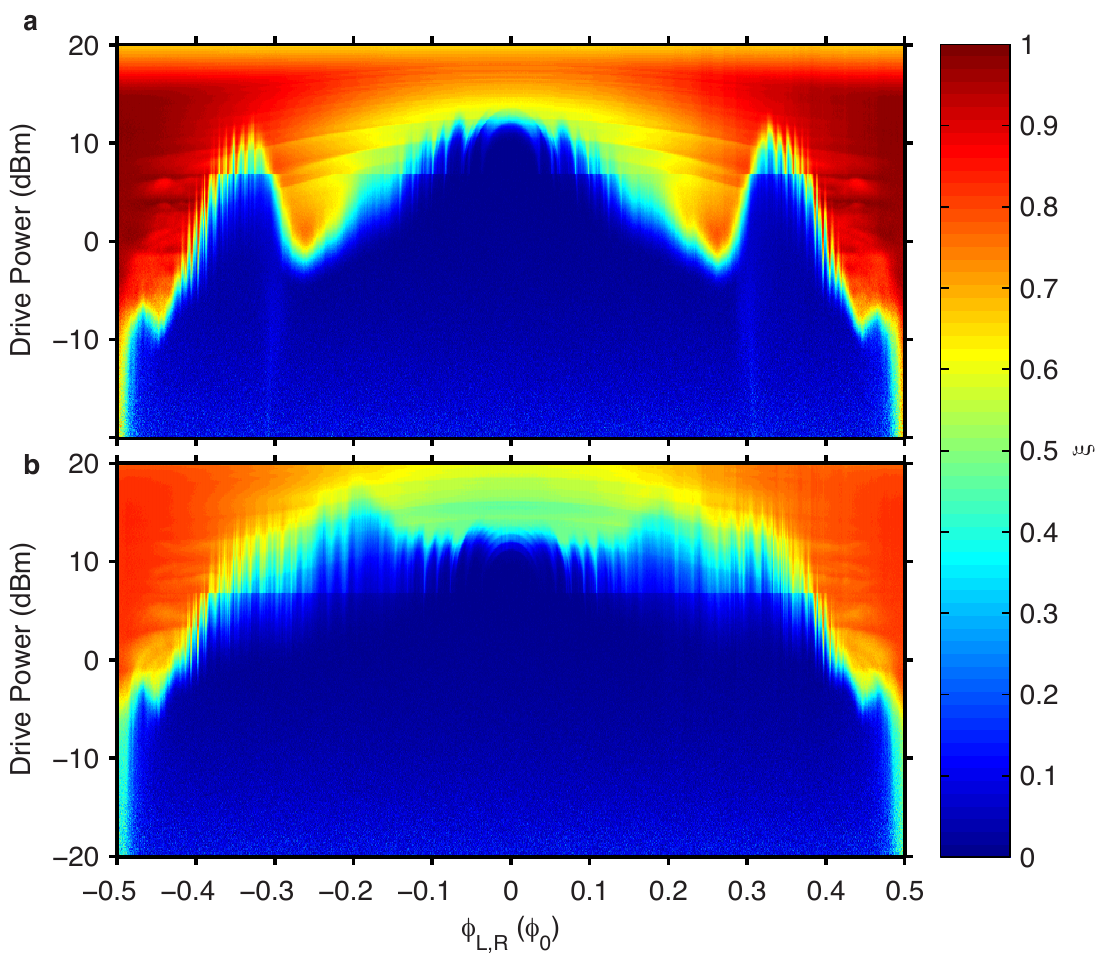}
\captionsetup{justification=raggedright,singlelinecheck=false,font=footnotesize}
\caption{\textbf{S4.}
Normalized dimer transmission $\xi$ as a function of drive power along the flux bias contour linearly moving both qubits from $(\Phi_L, \Phi_R)=(-1/2,-1/2)$ to $(\Phi_L, \Phi_R)=(1/2,1/2)$ for the antisymmetric (a) and symmetric (b) modes.
For each bias point power is swept and the transmission is recorded after $20$ seconds of averaging with an IF bandwidth of 500 kHz.
The long averaging time and high resolution reveal the fine structure of the transition from dark to bright state.
The horizontal line visible at 8 dBm is an artifact of the internal source configuration of the generator.
}
\end{figure}

\begin{figure}
\includegraphics[width=6in]{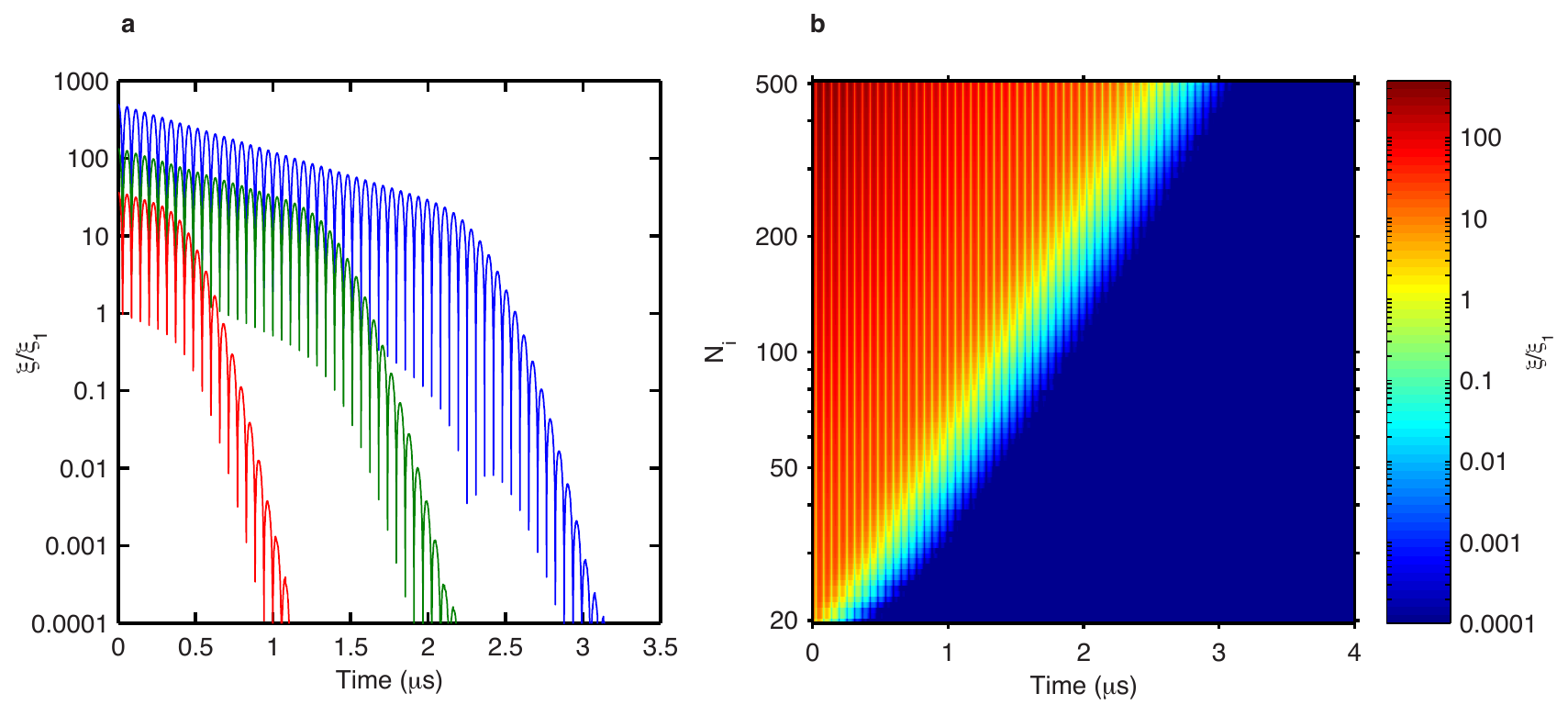}
\captionsetup{justification=raggedright,singlelinecheck=false,font=footnotesize}
\caption{
\textbf{S5. a,}
Toy model results demonstrating super-exponential decay behavior for $N_i=$ 500, 136, and 37 (top to bottom) with decay $\kappa=225$ kHz and $N_c=20$.
Assuming the existence of (1) a critical photon number $N_c$, (2) the dynamical behavior of the homodyne signal $\xi$ for the two different phases (linear oscillations for $N>N_c$ and no signal when localized), and (3) Poissonian photon decay, is sufficient to qualitatively reproduce the crossover signature observed in the experiment.
\textbf{b,} Analog of the phase diagram shown in figure (4c) reproduced using the toy model.
Further refinement of the toy model (not shown) by including the effects of critical slowing down as $N$ approaches $N_c$ causes super-exponential decay to occur earlier, but preserves the qualitative structure shown.
}
\end{figure}

\begin{figure}
\includegraphics[width=6in]{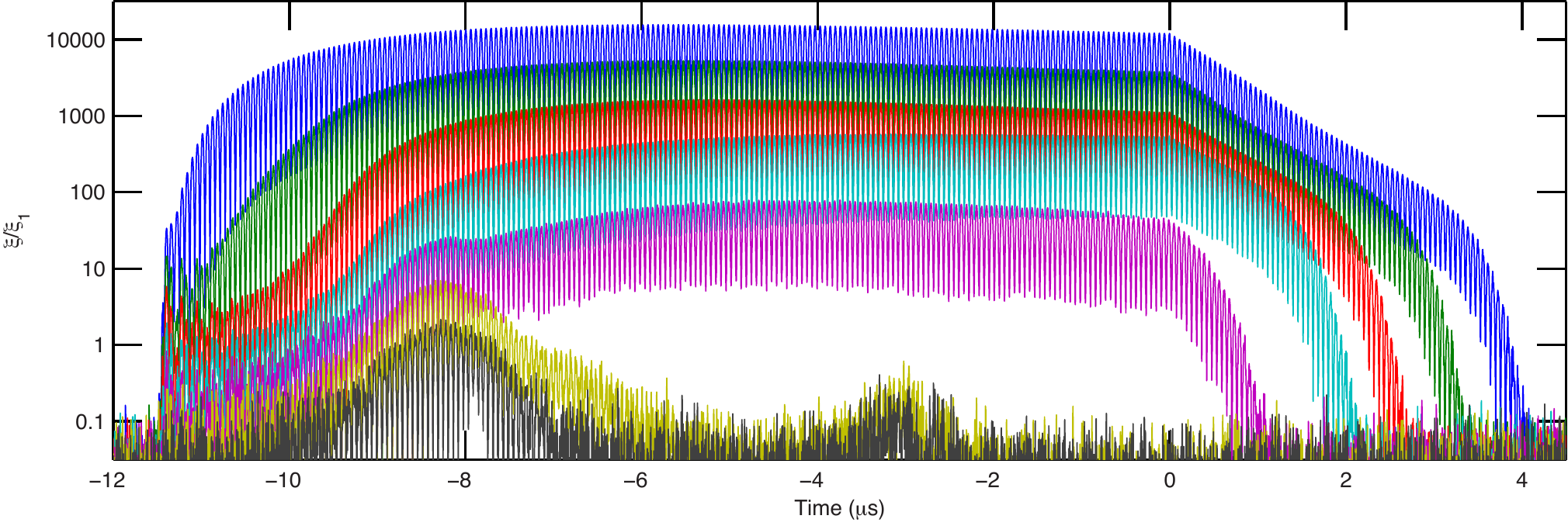}
\captionsetup{justification=raggedright,singlelinecheck=false,font=footnotesize}
\caption{
\textbf{S6.}
Driven dynamics using a long initialization drive.
Here $11.5 \mu s$ ($100/J$) initialization pulses are used with varying drive power (top to bottom 15 dBm to -15 dBm).
Drive ends and qubits are tuned into resonance at $t=0$.
The top five traces correspond to those shown in figure (4b) ($N_i \approx$ 12,000; 3,800; 1,100; 550; 40).
Lower power traces show the effects of finite photon-photon interactions even when the qubits are far detuned during the initialization procedure.
}
\end{figure}

\end{document}